\documentclass[12pt]{article}

\begin{document}

%%%%% The following lines create the SLAC Pub Title Page
%%
\thispagestyle{empty}
\renewcommand{\thefootnote}{\fnsymbol{footnote}}

%%%%% Substitute your Pub number, month and year in the following:
%%
%% \begin{flushright}
%% {\small
%% SLAC--PUB--XXX\\
%% Month Year\\}
%% \end{flushright}

\*\vspace{1.8cm}

%%%%% Title and Author Information:
%%
\begin{center}
{\bf\Large SOLITON GAS IN SPACE-CHARGE DOMINATED BEAMS}

\vspace{1cm}

{\bf\large Stephan I. TZENOV}\\
{\it\large Istituto Nazionale per la Fisica della Materia, Unit\`a
di Salerno, Via S. Allende, I-84081 Baronissi (Salerno) Italy}\\
\end{center}

\vfill

\begin{center}
{\bf\large
Abstract }
\end{center}

\begin{quote}
Based on the Vlasov-Maxwell equations describing the
self-consistent nonlinear beam dynamics and collective processes,
the evolution of an intense sheet beam propagating through a
periodic focusing field has been studied. It has been shown that
in the case of a beam with uniform phase space density the
Vlasov-Maxwell equations can be replaced exactly by the
hydrodynamic equations with a triple adiabatic pressure law
coupled to the Maxwell equations. We further demonstrate that
starting from the system of hydrodynamic and Maxwell equations a
set of coupled nonlinear Schrodinger equations for the slowly
varying amplitudes of density waves can be derived. In the case
where a parametric resonance between a certain mode of density
waves and the external focusing occurs, the slow evolution of the
resonant amplitudes in the cold-beam limit is shown to be governed
by a system of coupled Gross-Pitaevskii equations. Properties of
the nonlinear Schrodinger equation as well as properties of the
Gross-Pitaevskii equation are discussed, together with soliton and
condensate formation in intense particle beams.
\end{quote}

\vfill

%%%%%%%%%%%%%%%
%% Choose"Presented at," "Contributed to" for conference papers
%% or "Submitted to" for journal papers
%%%%%%%%%%%%%%%
%% \begin{center}
%% {\it (Invited talk presented at)}
%%   OR
%% {\it (Contributed to)}
%% (Text varies)
%% {\it Conference Name} (spell out completely) \\
%% {\it City, State, Country} (Location of Conference)\\
%% {\it Month Day--Month Day, Year}
%%   (Indicate duration of conference.)\\

%% OR\\

%% {\it Submitted to A Journal} (Spell out name of journal.)
%% \end{center}

\newpage
%%
%%%%% End of title page

%%%%% Following are the commands to create the rest of the SLAC Pub.
%%
%%%%% The next two lines change the line spacing to double space,
%%      if you should need to do that.
%%
%\renewcommand{\baselinestretch}{2}
%\normalsize

%%%%% Your paper starts here:
%%

%% To get page numbers in the rest of the paper:
%
\pagestyle{plain}

\renewcommand{\theequation}{\thesection.\arabic{equation}}

\setcounter{equation}{0}

\section{Introduction}

One of the main goals in the commissioning and operation of modern
high energy accelerators and storage rings is the achievement of
higher and higher beam currents and charge densities. That is why,
of particular importance are the effects of intense self-fields
due to space charge and current, influencing the beam propagation,
its stability and transport properties. In general, a complete
description of collective processes in intense charged particle
beams is provided by the Vlasov-Maxwell equations for the
self-consistent evolution of the beam distribution function and
the electromagnetic fields. Although the analytical basis (as
mentioned above) for modelling the dynamics and behaviour of
space-charge dominated beams is well established, a thorough and
satisfactory understanding of collective processes, detailed
equilibrium and formation of patterns and coherent structures is
far from being complete.

The present paper may be regarded as the first (to our knowledge)
attempt to take a view at the description of the evolution and the
collective behaviour of intense charged particle beams from an
entirely different perspective, as compared to the ones available
in the literature. We will be mainly interested in describing the
slow evolution of some coarse-grained quantities that are easily
measurable, such as the amplitudes of density waves. Due to the
nonlinear wave interaction contingent on the nonlinear coupling
between the Vlasov and Maxwell equations, one can expect a
formation of nontrivial coherent structure that might be fairly
stable in space and time. Here, we show that solitary wave
patterns in the beam density distribution are an irrevocable
feature, characteristic of intense beams. Moreover, density
condensates in the special case where a parametric resonance in
terms of wave frequency between a particular mode of the
fundamental density waves and the external focusing occurs, can be
formed.

The paper is organized as follows. It was previously shown
\cite{prstab} that in the case of a sheet beam with constant
phase-space density the Vlasov-Maxwell equations are fully
equivalent to a hydrodynamic model with zero heat flow and
triple-adiabatic equation-of-state. For consistency, in section 2,
we repeat the derivation from a slightly different perspective and
then use the hydrodynamic equations as a fundament for the
subsequent analysis. In section 3, we consider the case of a
smooth focusing where the time variation of the $\beta$-function
(respectively, the time variation of the density envelope
function) can be neglected. Further, we demonstrate that starting
from the hydrodynamic equations, and using the renormalization
group (RG) technique \cite{cgoono,nozaki,oono,shiwa} a system of
coupled nonlinear Schrodinger equations for the slowly varying
amplitudes of density waves can be derived. The purpose of section
4 is twofold. First, we study the case where a parametric
resonance between a particular mode of density waves and a
resonant Fourier harmonic of the external focusing (a resonant
harmonic from the ${\sqrt{\beta}}$-function Fourier decomposition)
occurs. It is shown that under certain conditions the evolution of
the resonant amplitudes of the forward and backward density waves
is governed by a system of coupled Gross-Pitaevskii equations (see
e.g. the review \cite{dalfovo} and the references therein).
Secondly, it is demonstrated that in the non-resonant case the
results obtained in the smooth focusing approximation coincide
with the ones for a periodic focusing up to second order in the
formal perturbation parameter. In section 5, the reduction of the
infinite system of coupled nonlinear Schrodinger equations to a
system of two coupled nonlinear Schrodinger equations is
discussed. Such reduction is possible if one neglects the effect
of all other modes and takes into account only the
self-interaction of a single mode with a particular wave number.
In the non-resonant case it is shown that the renormalized
solution for the beam density describes the process of formation
of {\it holes (cavitons)} in intense particle beams. The case
where a parametric resonance is present is more interesting. In
the cold-beam limit it is demonstrated that the evolution of the
forward and the backward resonant wave amplitudes can be well
described by a system of two coupled Gross-Pitaevskii equations.
Finally, conclusions are drawn in section 6.

\renewcommand{\theequation}{\thesection.\arabic{equation}}

\setcounter{equation}{0}

\section{Derivation of the Hydrodynamic Model}

We begin with the Hamiltonian describing the one-dimensional
betatron motion in the presence of space-charge field
\begin{equation}
{\widehat{H}} = {\frac {R} {2}} {\widehat{p}}^2 + {\frac
{G(\theta) {\widehat{x}}^2} {2R}} + V {\left( {\widehat{x}};
\theta \right)} + {\frac {eR} {E_s \beta_s^2}} \varphi_{sc}
{\left( {\widehat{x}}; \theta \right)}, \label{Haminit}
\end{equation}
\noindent where $e$ is the electron charge, $R$ is the mean radius
of the accelerator, $G(\theta)$ is the focusing strength of the
linear machine lattice, $V {\left( {\widehat{x}}; \theta \right)}$
is a contribution coming from nonlinear lattice elements
(sextupoles, octupoles, etc.) and $E_s$ and $\beta_s$ are the
energy and the relative velocity of the synchronous particle
respectively. In addition, $\varphi_{sc} {\left( {\widehat{x}};
\theta \right)}$ is the self-field potential due to space-charge,
which satisfies the one-dimensional Poisson equation
\begin{equation}
{\frac {\partial^2 \varphi_{sc}} {\partial {\widehat{x}}^2}} = -
{\frac {en} {\epsilon_0}} \int {\rm d} {\widehat{p}} f {\left(
{\widehat{x}}, {\widehat{p}}; \theta \right)}. \label{Poisinit}
\end{equation}
\noindent Here $n=N_p/V_t$ is the density of beam particles ($N_p$
is the number of particles in the beam, while $V_t$ is the volume
of the area occupied by the beam in the transverse direction),
$\epsilon_0$ is the dielectric susceptibility and $f {\left(
{\widehat{x}}, {\widehat{p}}; \theta \right)}$ is the distribution
function in phase space. Equations (\ref{Haminit}) and
(\ref{Poisinit}) can be written in an alternate form as
\begin{equation}
{\widehat{H}} = {\frac {R} {2}} {\widehat{p}}^2 + {\frac
{G(\theta) {\widehat{x}}^2} {2R}} + V {\left( {\widehat{x}};
\theta \right)} + \lambda {\widetilde{\varphi}} {\left(
{\widehat{x}}; \theta \right)}, \label{Hamalter}
\end{equation}
\begin{equation}
{\frac {\partial^2 {\widetilde{\varphi}}} {\partial
{\widehat{x}}^2}} = - \int {\rm d} {\widehat{p}} f {\left(
{\widehat{x}}, {\widehat{p}}; \theta \right)}, \label{Poisalter}
\end{equation}
\noindent where
\begin{equation}
\lambda = {\frac {R e^2 n} {\epsilon_0 E_s \beta_s^2}},
\label{Perveance}
\end{equation}
\noindent is the beam perveance, and $\varphi_{sc} = en
{\widetilde{\varphi}} / \epsilon_0$. Note that the beam perveance
$\lambda$ is a dimensionless quantity.

Next, we perform a canonical transformation
\begin{equation}
{\widehat{x}} = x {\sqrt{\beta}}, \qquad \qquad {\widehat{p}} =
{\frac {1} {\sqrt{\beta}}} (p - \alpha x), \label{Canonxp}
\end{equation}
\noindent defined by the generating function
\begin{equation}
F_2 {\left( {\widehat{x}}, p; \theta \right)} = {\frac
{{\widehat{x}} p} {\sqrt{\beta}}} - {\frac {\alpha
{\widehat{x}}^2} {2 \beta}}, \label{Genfunct}
\end{equation}
\noindent where $\alpha$ and $\beta$ (and $\gamma$) are the
well-known Twiss parameters satisfying the equations
\begin{equation}
{\frac {{\rm d} \alpha} {{\rm d} \theta}} = {\frac {G \beta} {R}}
- R \gamma, \qquad \qquad {\frac {{\rm d} \beta} {{\rm d} \theta}}
= - 2R \alpha, \qquad \qquad \beta \gamma - \alpha^2 = 1.
\label{Twisspar}
\end{equation}
\noindent As a result, we obtain the new Hamiltonian
\begin{equation}
H = {\frac {\dot{\chi}} {2}} {\left( p^2 + x^2 \right)} + V
{\left( x; \theta \right)} + \lambda {\sqrt{\beta}} U {\left( x;
\theta \right)}, \label{Hamstart}
\end{equation}
\noindent where the self-field potential $U {\left( x; \theta
\right)}$ ${\left( {\widetilde{\varphi}} = {\sqrt{\beta}} U
\right)}$ satisfies the equation
\begin{equation}
{\frac {\partial^2 U} {\partial x^2}} = - \int {\rm d} p f {\left(
x, p; \theta \right)}, \label{Poisstart}
\end{equation}
\noindent and
\begin{equation}
{\dot{\chi}} = {\frac {{\rm d} \chi} {{\rm d} \theta}} = {\frac
{R} {\beta}}, \label{Phaseadv}
\end{equation}
\noindent is the derivative of the phase advance with respect to
$\theta$.

We are now ready to write the Vlasov equation for the one-particle
distribution function $f {\left( x, p; \theta \right)}$ in the
two-dimensional phase space ${\left( x, p \right)}$. It reads as
\begin{equation}
{\frac {\partial f} {\partial \theta}} + {\dot{\chi}} p {\frac
{\partial f} {\partial x}} - {\left( {\dot{\chi}} x + {\frac
{\partial V} {\partial x}} + \lambda {\sqrt{\beta}} {\frac
{\partial U} {\partial x}} \right)} {\frac {\partial f} {\partial
p}} = 0, \label{Vlasoveq}
\end{equation}
\noindent and should be solved self-consistently with the Poisson
equation (\ref{Poisstart}). Following Davidson et al.
\cite{prstab}, we consider the case where the distribution
function $f {\left( x, p; \theta \right)}$ is constant
(independent of $x$, $p$ and $\theta$) inside a region in phase
space confined by the simply connected boundary curves $p_{(+)}
{\left( x; \theta \right)}$ and $p_{(-)} {\left( x; \theta
\right)}$, and zero outside. In other words,
\begin{equation}
f {\left( x, p; \theta \right)} = {\cal C}, \qquad {\rm for}
\qquad p_{(-)} {\left( x; \theta \right)} < p < p_{(+)} {\left( x;
\theta \right)}, \label{Distfun}
\end{equation}
\noindent and $f {\left( x, p; \theta \right)} = 0$ otherwise. The
proof that such solution to the Vlasov equation (\ref{Vlasoveq})
exists can be found in appendix A.

The equations for the boundary curves $p_{(\pm)} {\left( x; \theta
\right)}$ can be derived as follows. We substitute the explicit
form
\begin{equation}
f {\left( x, p; \theta \right)} = {\cal C} {\left[ {\cal H}
{\left( p - p_{(-)} {\left( x; \theta \right)} \right)} - {\cal H}
{\left( p - p_{(+)} {\left( x; \theta \right)} \right)} \right]}
\label{Dfunexform}
\end{equation}
\noindent of the distribution function into the Vlasov equation
(\ref{Vlasoveq}). Here ${\cal H} (z)$ is the well-known Heaviside
function. Using the fact that the derivative of the Heaviside
function with respect to its argument equals the Dirac
$\delta$-function, we obtain
\begin{eqnarray}
- \delta {\left( p - p_{(-)} \right)} {\frac {\partial p_{(-)}}
{\partial \theta}} + \delta {\left( p - p_{(+)} \right)} {\frac
{\partial p_{(+)}} {\partial \theta}} + {\dot{\chi}} p {\left[ -
\delta {\left( p - p_{(-)} \right)} {\frac {\partial p_{(-)}}
{\partial x}} + \delta {\left( p - p_{(+)} \right)} {\frac
{\partial p_{(+)}} {\partial x}} \right]} \nonumber
\end{eqnarray}
\begin{equation}
- {\left( {\dot{\chi}} x + {\frac {\partial V} {\partial x}} +
\lambda {\sqrt{\beta}} {\frac {\partial U} {\partial x}} \right)}
{\left[ \delta {\left( p - p_{(-)} \right)} - \delta {\left( p -
p_{(+)} \right)} \right]} = 0. \label{Basderiv}
\end{equation}
\noindent Multiplying equation (\ref{Basderiv}) first by $1$ and
then by $p$, and integrating the result over $p$, we readily
obtain
\begin{equation}
{\frac {\partial} {\partial \theta}} {\left( p_{(+)} - p_{(-)}
\right)} + {\frac {\dot{\chi}} {2}} {\frac {\partial} {\partial
x}} {\left( p_{(+)}^2 - p_{(-)}^2 \right)} = 0, \label{Firstmom}
\end{equation}
\begin{equation}
{\frac {1} {2}} {\frac {\partial} {\partial \theta}} {\left(
p_{(+)}^2 - p_{(-)}^2 \right)} + {\frac {\dot{\chi}} {3}} {\frac
{\partial} {\partial x}} {\left( p_{(+)}^3 - p_{(-)}^3 \right)} =
- {\left( p_{(+)} - p_{(-)} \right)} {\left( {\dot{\chi}} x +
{\frac {\partial V} {\partial x}} + \lambda {\sqrt{\beta}} {\frac
{\partial U} {\partial x}} \right)}, \label{Secondmom}
\end{equation}
\begin{equation}
{\frac {\partial^2 U} {\partial x^2}} = - {\cal C} {\left( p_{(+)}
- p_{(-)} \right)}, \label{Poissonmom}
\end{equation}

It is convenient to recast equations (\ref{Firstmom}) -
(\ref{Poissonmom}) in a more familiar form, widely used in
hydrodynamics. Let us define
\begin{equation}
\rho = \int \limits_{- \infty}^{\infty} {\rm d} p f {\left( x, p;
\theta \right)} = {\cal C} {\left( p_{(+)} - p_{(-)} \right)},
\label{Defdens}
\end{equation}
\begin{equation}
\rho v = \int \limits_{- \infty}^{\infty} {\rm d} p p f {\left( x,
p; \theta \right)} = {\frac {\cal C} {2}} {\left( p_{(+)}^2 -
p_{(-)}^2 \right)}, \label{Defveloc}
\end{equation}
\noindent where $\rho {\left( x; \theta \right)}$ and $v {\left(
x; \theta \right)}$ are the density and the current velocity,
respectively. From the two definitions above, it follows that
\begin{equation}
v = {\frac {1} {2}} {\left( p_{(+)} + p_{(-)} \right)}.
\label{Velocpp}
\end{equation}
\noindent Defining further the pressure ${\cal P} {\left( x;
\theta \right)}$ and the heat flow ${\cal Q} {\left( x; \theta
\right)}$ and using equation (\ref{Velocpp}), we have
\begin{equation}
{\cal P} = \int \limits_{- \infty}^{\infty} {\rm d} p (p-v)^2 f
{\left( x, p; \theta \right)} = {\frac {1} {12 {\cal C}^2}}
{\left[ {\cal C} {\left( p_{(+)} - p_{(-)} \right)} \right]}^3,
\label{Defpress}
\end{equation}
\begin{equation}
{\cal Q} = \int \limits_{- \infty}^{\infty} {\rm d} p (p-v)^3 f
{\left( x, p; \theta \right)} = 0. \label{Defheatfl}
\end{equation}
\noindent From equation (\ref{Defdens}), it follows that the
pressure can be expressed as
\begin{equation}
{\cal P} = {\frac {{\cal P}_0} {{\widehat{\rho}}_0^3}} \rho^3,
\qquad \qquad {\frac {{\cal P}_0} {{\widehat{\rho}}_0^3}} = {\frac
{1} {12 {\cal C}^2}}. \label{Pressexp}
\end{equation}
\noindent In addition, it is straightforward to verify that
\begin{equation}
{\frac {\cal C} {3}} {\left( p_{(+)}^3 - p_{(-)}^3 \right)} =
{\cal P} + \rho v^2, \label{Thirdmom}
\end{equation}
\noindent which provides a closure for the equations
(\ref{Firstmom}) and (\ref{Secondmom}) governing the evolution of
the moments (boundary curves). With all the above definitions and
relations in hand, we can write the system of hydrodynamic
equations (supplemented by the Poisson equation) in the form
\begin{equation}
{\frac {\partial \rho} {\partial \theta}} + {\dot{\chi}} {\frac
{\partial} {\partial x}} (\rho v) = 0, \label{Densityeq}
\end{equation}
\begin{equation}
{\frac {\partial} {\partial \theta}} (\rho v) + {\dot{\chi}}
{\frac {\partial} {\partial x}} {\left( {\cal P} + \rho v^2
\right)} = - \rho {\left( {\dot{\chi}} x + {\frac {\partial V}
{\partial x}} + \lambda {\sqrt{\beta}} {\frac {\partial U}
{\partial x}} \right)}, \label{Velocityeq}
\end{equation}
\begin{equation}
{\frac {\partial^2 U} {\partial x^2}} = - \rho. \label{Poissoneq}
\end{equation}
\noindent Using the continuity equation (\ref{Densityeq}), we cast
the system of equations (\ref{Densityeq}) - (\ref{Poissoneq}) in
its final form
\begin{equation}
{\frac {\partial \rho} {\partial \tau}} + {\frac {\partial}
{\partial x}} (\rho v) = 0, \label{Continuityeq}
\end{equation}
\begin{equation}
{\frac {\partial v} {\partial \tau}} + v {\frac {\partial v}
{\partial x}} + v_T^2 {\frac {\partial} {\partial x}} {\left(
\rho^2 \right)} = - x - {\frac {\beta} {R}} {\frac {\partial V}
{\partial x}} - \lambda {\frac {\beta {\sqrt{\beta}}} {R}} {\frac
{\partial U} {\partial x}}, \label{Cvelocityeq}
\end{equation}
\begin{equation}
{\frac {\partial^2 U} {\partial x^2}} = - \rho, \label{Poissoeq}
\end{equation}
\noindent which will be the starting point for the subsequent
analysis. Here
\begin{equation}
v_T^2 = {\frac {3 {\cal P}_0} {2 {\widehat{\rho}}_0^3}},
\label{Thveloc}
\end{equation}
\noindent is the normalized thermal speed-squared, and $\tau =
\chi (\theta) + \tau_0$ is the new independent "time"-variable.

\renewcommand{\theequation}{\thesection.\arabic{equation}}

\setcounter{equation}{0}

\section{Renormalization Group Reduction of the Hydrodynamic Equations}

Before proceeding further, we make an important remark. Let
$F(\theta)$ be a periodic function of $\theta$ with period $2
\pi$. Noting that the phase advance $\chi (\theta)$ can be
represented in the form
\begin{equation}
\chi (\theta) = \nu \theta + \chi_p (\theta) - \chi_p (\pi),
\label{Phaseadva}
\end{equation}
\noindent where $\nu$ is the betatron tune, and $\chi_p (\theta +
2 \pi) = \chi_p (\theta)$, we can expand $F(\theta)$ regarded as a
function of $\chi$ (respectively $\tau$) in a Fourier series in
the new variable $\chi$ (respectively $\tau$) as follows
\begin{equation}
F(\theta) = \sum \limits_{n = - \infty}^{\infty} {\cal A}_n \exp
{\left( in {\frac {\chi} {\nu}} \right)}. \label{Fourchi}
\end{equation}
\noindent The expansion coefficients ${\cal A}_n$ are given by the
expressions
\begin{equation}
{\cal A}_n = {\frac {1} {2 \pi \nu}} \int \limits_{- \pi \nu}^{\pi
\nu} {\rm d} \chi F(\theta) \exp {\left( -in {\frac {\chi} {\nu}}
\right)}. \label{Expancoeff}
\end{equation}
\noindent Using the definition of the phase advance
(\ref{Phaseadva}), and choosing $\tau_0 = \chi_p (\pi)$, we can
rewrite the Fourier expansion (\ref{Fourchi}) as
\begin{equation}
F(\theta) = \sum \limits_{n = - \infty}^{\infty} {\cal B}_n \exp
{\left( in {\frac {\tau} {\nu}} \right)}, \label{Fourtau}
\end{equation}
\noindent where the expansion coefficients ${\cal B}_n$ are given
by the expressions
\begin{equation}
{\cal B}_n = {\frac {R} {2 \pi \nu}} \int \limits_{- \pi}^{\pi}
{\rm d} \theta {\frac {F(\theta)} {\beta (\theta)}} e^{-in \theta}
\exp {\left[ -in {\frac {\chi_p (\theta)} {\nu}} \right]}.
\label{Expancoefb}
\end{equation}

In what follows, we consider the case where the external potential
$V$ is zero, so that the set of equations to be analyzed acquires
the form
\begin{equation}
{\frac {\partial \rho} {\partial \tau}} + {\frac {\partial}
{\partial x}} (\rho v) = 0, \label{Contineq}
\end{equation}
\begin{equation}
{\frac {\partial v} {\partial \tau}} + v {\frac {\partial v}
{\partial x}} + v_T^2 {\frac {\partial} {\partial x}} {\left(
\rho^2 \right)} = - x - \lambda g(\tau) {\frac {\partial U}
{\partial x}}, \label{Cveloceq}
\end{equation}
\begin{equation}
{\frac {\partial^2 U} {\partial x^2}} = - \rho, \label{Poiseq}
\end{equation}
\noindent where
\begin{equation}
g(\tau) = \sum \limits_{n = - \infty}^{\infty} a_n \exp {\left( in
{\frac {\tau} {\nu}} \right)}, \label{Functg}
\end{equation}
\begin{equation}
a_n = {\frac {1} {2 \pi \nu}} \int \limits_{- \pi}^{\pi} {\rm d}
\theta {\sqrt{\beta (\theta)}} e^{-in \theta} \exp {\left[ -in
{\frac {\chi_p (\theta)} {\nu}} \right]}. \label{Excoefb}
\end{equation}

Further, we assume that there exists a nontrivial solution to
equations (\ref{Contineq}) - (\ref{Poiseq}) in the interval $x \in
{\left( - x_{(-)}, x_{(+)} \right)}$, and that the sheet beam
density is zero ${\left( \varrho = 0 \right)}$ outside of the
interval. We introduce also the ansatz
\begin{equation}
\rho {\left( x; \tau \right)} = {\frac {1} {{\cal E}}} + \epsilon
R {\left( x; \tau \right)}, \qquad v {\left( x; \tau \right)} =
{\frac {x} {{\cal E}}} {\frac {{\rm d} {\cal E}} {{\rm d} \tau}} +
\epsilon u {\left( x; \tau \right)}, \qquad U {\left( x; \tau
\right)} = - {\frac {x^2} {2 {\cal E}}} + \epsilon {\cal U}
{\left( x; \tau \right)}, \label{Ansatzb}
\end{equation}
\noindent where the envelope function ${\cal E} (\tau)$ is a
solution to the equation
\begin{equation}
{\frac {{\rm d}^2 {\cal E}} {{\rm d} \tau^2}} + {\cal E} = \lambda
g(\tau), \label{Enveleq}
\end{equation}
\noindent and therefore can be represented in explicit form as
\begin{equation}
{\cal E} (\tau) = \lambda a_0 + \lambda \sum \limits_{n \neq 0}
b_n e^{in \tau / \nu}, \qquad \qquad b_n = {\frac {a_n} {1 - n^2 /
\nu^2}}. \label{Fourenvel}
\end{equation}
\noindent It enables us to rewrite equations (\ref{Contineq}) -
(\ref{Poiseq}) as follows
\begin{equation}
{\frac {\partial R} {\partial \tau}} + {\frac {1} {{\cal E}}}
{\frac {\partial u} {\partial x}} + {\frac {{\dot{\cal E}}} {{\cal
E}}} {\frac {\partial} {\partial x}} (xR) + \epsilon {\frac
{\partial} {\partial x}} {\left( R u \right)} = 0,
\label{Contineqr}
\end{equation}
\begin{equation}
{\frac {\partial u} {\partial \tau}} + {\frac {{\dot{\cal E}}}
{{\cal E}}} {\frac {\partial} {\partial x}} (xu) + {\frac {2
v_T^2} {{\cal E}}}  {\frac {\partial R} {\partial x}} + \epsilon u
{\frac {\partial u} {\partial x}} + \epsilon v_T^2 {\frac
{\partial} {\partial x}} {\left( R^2 \right)} =  - \lambda g
{\frac {\partial {\cal U}} {\partial x}}, \label{Hydrodyn}
\end{equation}
\begin{eqnarray}
{\frac {\partial^2 {\cal U}} {\partial x^2}} = - R. \nonumber
\end{eqnarray}
\noindent Next, we differentiate the first equation
(\ref{Contineqr}) of the above system with respect to $\tau$ and
the second equation (\ref{Hydrodyn}) with respect to $x$. After
summing these up, we obtain
\begin{eqnarray}
{\cal E} {\frac {\partial} {\partial \tau}} {\left( {\cal E}
{\frac {\partial R} {\partial \tau}} \right)} - 2 v_T^2 {\frac
{\partial^2 R} {\partial x^2}} + \lambda g {\cal E} R \nonumber
\end{eqnarray}
\begin{equation}
= {\dot{\cal E}} {\frac {\partial^2} {\partial x^2}} (xu) - {\cal
E} {\frac {\partial} {\partial \tau}} {\left[ {\dot{\cal E}}
{\frac {\partial} {\partial x}} (xR) \right]} + \epsilon {\cal E}
{\frac {\partial^2} {\partial x^2}} {\left( {\frac {u^2} {2}} +
v_T^2 R^2 \right)} - \epsilon {\cal E} {\frac {\partial} {\partial
\tau}} {\left[ {\cal E} {\frac {\partial} {\partial x}} (Ru)
\right]}. \label{Basic1}
\end{equation}
\noindent Equation (\ref{Basic1}) supplemented with equation
(\ref{Contineqr}) rewritten as
\begin{equation}
{\cal E} {\frac {\partial R} {\partial \tau}} + {\frac {\partial
u} {\partial x}} + {\dot{\cal E}} {\frac {\partial} {\partial x}}
(xR) + \epsilon {\cal E} {\frac {\partial} {\partial x}} {\left( R
u \right)} = 0, \label{Basic2}
\end{equation}
\noindent comprises the basic system of equations for the analysis
in the subsequent exposition.

In what follows, we consider the case of smooth focusing, where
the time variation of $g(\tau)$ [${\cal E} (\tau)$, respectively]
can be neglected. In the next section, we will show that when the
frequencies of all fundamental modes are sufficiently far from a
parametric resonance, this assumption holds true to second order
in the perturbation parameter $\epsilon$ even in the case when
such time variation is present. Thus, the basic equations
(\ref{Basic1}) and (\ref{Basic2}) acquire the form
\begin{equation}
{\cal E}_0^2 {\frac {\partial^2 R} {\partial \tau^2}} - 2 v_T^2
{\frac {\partial^2 R} {\partial x^2}} + {\cal E}_0^2 R  = \epsilon
{\cal E}_0 {\frac {\partial^2} {\partial x^2}} {\left( {\frac
{u^2} {2}} + v_T^2 R^2 \right)} - \epsilon {\cal E}_0^2 {\frac
{\partial^2} {\partial \tau \partial x}} (Ru), \label{Basic11}
\end{equation}
\begin{equation}
{\cal E}_0 {\frac {\partial R} {\partial \tau}} + {\frac {\partial
u} {\partial x}} + \epsilon {\cal E}_0 {\frac {\partial} {\partial
x}} {\left( R u \right)} = 0, \label{Basic22}
\end{equation}
\noindent where
\begin{equation}
{\cal E}_0 = \lambda a_0. \label{Ezerodef}
\end{equation}
\noindent Let us further assume that the actual dependence of $R$
and $u$ on the independent variables is given by
\begin{equation}
R = R {\left( x, \xi; \tau \right)}, \qquad \qquad u = u {\left(
x, \xi; \tau \right)}, \label{Actdepen}
\end{equation}
\noindent where
\begin{equation}
\xi = \epsilon x \label{Xivar}
\end{equation}
\noindent is a slow spatial variable. Therefore, the basic
equations (\ref{Basic11}) and (\ref{Basic22}) can be rewritten as
\begin{eqnarray}
{\cal E}_0^2 {\frac {\partial^2 R} {\partial \tau^2}} - 2 v_T^2
{\left( {\frac {\partial^2} {\partial x^2}} + 2 \epsilon {\frac
{\partial^2} {\partial x \partial \xi}} + \epsilon^2 {\frac
{\partial^2} {\partial \xi^2}} \right)} R + {\cal E}_0^2 R
\nonumber
\end{eqnarray}
\begin{equation}
= \epsilon {\cal E}_0 {\left( {\frac {\partial^2} {\partial x^2}}
+ 2 \epsilon {\frac {\partial^2} {\partial x \partial \xi}} +
\epsilon^2 {\frac {\partial^2} {\partial \xi^2}} \right)} {\left(
{\frac {u^2} {2}} + v_T^2 R^2 \right)} - \epsilon {\cal E}_0^2
{\left( {\frac {\partial^2} {\partial \tau
\partial x}} + \epsilon {\frac {\partial^2} {\partial \tau
\partial \xi}} \right)} (Ru), \label{Basic12}
\end{equation}
\begin{equation}
{\cal E}_0 {\frac {\partial R} {\partial \tau}} + {\frac {\partial
u} {\partial x}} + \epsilon {\frac {\partial u} {\partial \xi}} +
\epsilon {\cal E}_0 {\left( {\frac {\partial} {\partial x}} +
\epsilon {\frac {\partial} {\partial \xi}} \right)} {\left( R u
\right)} = 0, \label{Basic21}
\end{equation}

Following the basic idea of the RG method, we represent the
solution to equations (\ref{Basic12}) and (\ref{Basic21}) in the
form of a standard perturbation expansion \cite{nayfeh} in the
formal small parameter $\epsilon$ as
\begin{equation}
R = \sum \limits_{k=0}^{\infty} \epsilon^k R_k, \qquad \qquad u =
\sum \limits_{k=0}^{\infty} \epsilon^k u_k, \label{Naiverhov}
\end{equation}
\noindent The zeroth-order equations (\ref{Basic12}) and
(\ref{Basic21}) read as
\begin{equation}
\lambda^2 a_0^2 {\frac {\partial^2 R_0} {\partial \tau^2}} - 2
v_T^2 {\frac {\partial^2 R_0} {\partial x^2}} + \lambda^2 a_0^2
R_0 = 0, \label{Zero1}
\end{equation}
\begin{equation}
\lambda a_0 {\frac {\partial R_0} {\partial \tau}} + {\frac
{\partial u_0} {\partial x}} = 0. \label{Zero2}
\end{equation}
\noindent Their solutions can be found in a straightforward manner
to be
\begin{equation}
R_0 {\left( x, \xi; \tau \right)} = \sum \limits_{m \neq 0}
{\left[ A_m (\xi) e^{i z_m^{(+)}} + B_m (\xi) e^{i z_m^{(-)}}
\right]}, \label{Solr0}
\end{equation}
\begin{equation}
u_0 {\left( x, \xi; \tau \right)} = - {\frac {2 \pi} {\sigma^2}}
\sum \limits_{m \neq 0} {\frac {\omega_m} {m}} {\left[ A_m (\xi)
e^{i z_m^{(+)}} - B_m (\xi) e^{i z_m^{(-)}} \right]}.
\label{Solu0}
\end{equation}
\noindent Here $A_m$ and $B_m$ are constant complex amplitudes (to
this end dependent on the slow variable $\xi$ only) of the
backward and the forward wave solution to equation (\ref{Zero1}),
respectively. These will be the subject of renormalization at the
final step of the renormalization group procedure resulting in RG
equations governing their slow evolution. Furthermore,
\begin{equation}
z_m^{(\pm)} {\left( x; \tau \right)} = \omega_m \tau \pm m \sigma
x, \qquad \qquad \sigma = {\frac {2 \pi} {x_{(+)} + x_{(-)}}},
\qquad \qquad \lambda a_0 = {\frac {2 \pi} {\sigma}}.
\label{Defhisig}
\end{equation}
\noindent The discrete mode frequencies $\omega_m$ are determined
from the dispersion relation
\begin{equation}
\omega_m^2 = 1 + K^2 m^2, \qquad \qquad K = {\frac {v_T^2
\sigma^4} {2 \pi^2}}. \label{Dispersion}
\end{equation}
\noindent It can be easily verified that the above choice of
parameters leads to
\begin{equation}
\int \limits_{-x_{(-)}}^{x_{(+)}} {\rm d} x R_0 {\left( x; \theta
\right)} = 0, \label{Conservation}
\end{equation}
\noindent which means that linear perturbation to the uniform
density $\rho_0 = {\cal E}_0^{-1}$ average to zero and do not
affect the normalization properties on the interval ${\left(
x^{(-)}, x^{(+)} \right)}$. In addition, the following conventions
and notations
\begin{equation}
\omega_{-m} = - \omega_m, \quad \quad A_{-m} = A_m^{\ast}, \quad
\quad B_{-m} = B_m^{\ast} \label{Notations}
\end{equation}
\noindent have been introduced.

The first order perturbation equation for $R_1$ reads as
\begin{eqnarray}
{\frac {\partial^2 R_1} {\partial \tau^2}} - {\frac {K^2}
{\sigma^2}} {\frac {\partial^2 R_1} {\partial x^2}} + R_1 = {\frac
{2iK^2} {\sigma}} \sum \limits_{m \neq 0} m {\left( {\frac
{\partial A_m} {\partial \xi}} e^{i z_m^{(+)}} - {\frac {\partial
B_m} {\partial \xi}} e^{i z_m^{(-)}} \right)} \nonumber
\end{eqnarray}
\begin{equation}
- {\frac {\pi} {\sigma}} \sum \limits_{m, n \neq 0} {\left[
\gamma_{mn}^{(+)} A_m A_n e^{i {\left( z_m^{(+)} + z_n^{(+)}
\right)}} + 2 \gamma_{mn}^{(-)} A_m B_n e^{i {\left( z_m^{(+)} +
z_n^{(-)} \right)}} + \gamma_{mn}^{(+)} B_m B_n e^{i {\left(
z_m^{(-)} + z_n^{(-)} \right)}} \right]}, \label{Firstr1}
\end{equation}
\noindent where
\begin{equation}
\gamma_{mn}^{(\pm)} = (m \pm n) {\left[ {\left( \omega_m +
\omega_n \right)} {\left( {\frac {\omega_m} {m}} \pm {\frac
{\omega_n} {n}} \right)} + (m \pm n) {\left( K^2 \pm {\frac
{\omega_m \omega_n} {mn}} \right)} \right]}. \label{Gammacoeff}
\end{equation}
\noindent The solution for $R_1$ is readily obtained to be
\begin{eqnarray}
R_1 {\left( x, \xi; \tau \right)} = {\frac {K^2 \tau} {\sigma}}
\sum \limits_{m \neq 0} {\frac {m} {\omega_m}} {\left( {\frac
{\partial A_m} {\partial \xi}} e^{i z_m^{(+)}} - {\frac {\partial
B_m} {\partial \xi}} e^{i z_m^{(-)}} \right)} \nonumber
\end{eqnarray}
\begin{equation}
- {\frac {\pi} {\sigma}} \sum \limits_{m, n \neq 0} {\left[
\alpha_{mn}^{(+)} A_m A_n e^{i {\left( z_m^{(+)} + z_n^{(+)}
\right)}} + 2 \alpha_{mn}^{(-)} A_m B_n e^{i {\left( z_m^{(+)} +
z_n^{(-)} \right)}} + \alpha_{mn}^{(+)} B_m B_n e^{i {\left(
z_m^{(-)} + z_n^{(-)} \right)}} \right]}, \label{Firsolr1}
\end{equation}
\noindent where
\begin{equation}
\alpha_{mn}^{(\pm)} = {\frac {\gamma_{mn}^{(\pm)}} {{\cal
D}_{mn}^{(\pm)}}}, \label{Alphacoeff}
\end{equation}
\begin{equation}
{\cal D}_{mn}^{(\pm)} = 1 - {\left( \omega_m + \omega_n \right)}^2
+ K^2 (m \pm n)^2. \label{Discriminant}
\end{equation}
\noindent Note that the (infinite dimensional) matrices
${\widehat{\alpha}}^{(\pm)}$ are symmetric, i.e.
\begin{equation}
\alpha_{mn}^{(\pm)} = \alpha_{nm}^{(\pm)}, \qquad \qquad
\alpha_{m, \mp m}^{(\pm)} = 0. \label{Symmetmatr}
\end{equation}
\noindent Having determined $R_1 {\left( x; \tau \right)}$, the
first-order current velocity $u_1 {\left( x; \tau \right)}$ can be
found in a straightforward manner. The result is
\begin{eqnarray}
u_1 {\left( x, \xi; \tau \right)} = - {\frac {2 \pi K^2 \tau}
{\sigma^3}} \sum \limits_{m \neq 0} {\left( {\frac {\partial A_m}
{\partial \xi}} e^{i z_m^{(+)}} + {\frac {\partial B_m} {\partial
\xi}} e^{i z_m^{(-)}} \right)} \nonumber
\end{eqnarray}
\begin{eqnarray}
- {\frac {2 \pi i} {\sigma^3}} \sum \limits_{m \neq 0} {\frac {1}
{m^2 \omega_m}} {\left( {\frac {\partial A_m} {\partial \xi}} e^{i
z_m^{(+)}} + {\frac {\partial B_m} {\partial \xi}} e^{i z_m^{(-)}}
\right)} \nonumber
\end{eqnarray}
\begin{equation}
+ {\frac {2 \pi^2} {\sigma^3}} \sum \limits_{m, n \neq 0} {\left[
\beta_{mn}^{(+)} A_m A_n e^{i {\left( z_m^{(+)} + z_n^{(+)}
\right)}} + 2 \beta_{mn}^{(-)} A_m B_n e^{i {\left( z_m^{(+)} +
z_n^{(-)} \right)}} - \beta_{mn}^{(+)} B_m B_n e^{i {\left(
z_m^{(-)} + z_n^{(-)} \right)}} \right]}, \label{Firsolu1}
\end{equation}
\noindent where
\begin{equation}
\beta_{mn}^{(\pm)} = {\frac {\omega_m} {m}} \pm {\frac {\omega_n}
{n}} + \alpha_{mn}^{(\pm)} {\frac {\omega_m + \omega_n} {m \pm
n}}, \qquad \qquad \beta_{m, -m}^{(+)} = 0. \label{Betacoeff}
\end{equation}
\noindent Note that the (infinite dimensional) matrix
${\widehat{\beta}}^{(+)}$ possesses the same symmetry properties
as those displayed by equation (\ref{Symmetmatr}) possessed by the
matrix ${\widehat{\alpha}}^{(+)}$, while ${\widehat{\beta}}^{(-)}$
is antisymmetric (and evidently $\beta_{mm}^{(-)} = 0$).

In second order, the equation for $R_2 (x, \xi; \tau)$ acquires a
form similar to that of equation (\ref{Firstr1}). It is important
to emphasize that this assertion holds true in every subsequent
order. Each entry on the right-hand-sides of the corresponding
equations can be calculated explicitly utilizing the already
determined quantities from the previous orders. The
right-hand-side of the equation for $R_2 (x, \xi; \tau)$ contains
terms which yield oscillating contributions with constant
amplitudes to the solution for $R_2 (x, \xi; \tau)$. Apart from
these, there are resonant terms (proportional to ${\rm e}^{i
z_m^{(\pm)} {\left( x; \tau \right)}}$) leading to a secular
contribution. To complete the renormalization group reduction of
the hydrodynamic equations, we select these particular resonant
second-order terms on the right-hand-side of the equation
determining $R_2 (x, \xi; \tau)$. The latter can be written as
\begin{eqnarray}
{\frac {\partial^2 R_2} {\partial \tau^2}} - {\frac {K^2}
{\sigma^2}} {\frac {\partial^2 R_2} {\partial x^2}} + R_2 = {\frac
{K^2} {\sigma^2}} \sum \limits_{m \neq 0} {\left( {\frac
{\partial^2 A_m} {\partial \xi^2}} e^{i z_m^{(+)}} + {\frac
{\partial^2 B_m} {\partial \xi^2}} e^{i z_m^{(-)}} \right)}
\nonumber
\end{eqnarray}
\begin{eqnarray}
+ {\frac {2iK^4 \tau} {\sigma^2}} \sum \limits_{m \neq 0} {\frac
{m^2} {\omega_m}} {\left( {\frac {\partial^2 A_m} {\partial
\xi^2}} e^{i z_m^{(+)}} + {\frac {\partial^2 B_m} {\partial
\xi^2}} e^{i z_m^{(-)}} \right)} \nonumber
\end{eqnarray}
\begin{equation}
+ {\frac {4 \pi^2} {\sigma^2}} \sum \limits_{m, n \neq 0} {\left[
{\left( \Gamma_{mn}^{(+)} {\left| A_n \right|}^2 +
\Gamma_{mn}^{(-)} {\left| B_n \right|}^2 \right)} A_m e^{i
z_m^{(+)}} + {\left( \Gamma_{mn}^{(-)} {\left| A_n \right|}^2 +
\Gamma_{mn}^{(+)} {\left| B_n \right|}^2 \right)} B_m e^{i
z_m^{(-)}} \right]}, \label{Secondr1}
\end{equation}
\noindent where
\begin{equation}
\Gamma_{mn}^{(+)} = m^2 {\left[ \beta_{mn}^{(+)} {\left( {\frac
{\omega_m} {m}} + {\frac {\omega_n} {n}} \right)} +
\alpha_{mn}^{(+)} {\left( K^2 + {\frac {\omega_m \omega_n} {mn}}
\right)} \right]} {\left( 1 - {\frac {\delta_{mn}} {2}} \right)},
\label{Gammamnp}
\end{equation}
\begin{equation}
\Gamma_{mn}^{(-)} = m^2 {\left[ \beta_{mn}^{(-)} {\left( {\frac
{\omega_m} {m}} - {\frac {\omega_n} {n}} \right)} +
\alpha_{mn}^{(-)} {\left( K^2 - {\frac {\omega_m \omega_n} {mn}}
\right)} \right]}, \label{Gammamnm}
\end{equation}
\noindent Some straightforward algebra yields the solution for
$R_2 {\left( x, \xi; \tau \right)}$ in the form
\begin{eqnarray}
R_2 {\left( x, \xi; \tau \right)} = {\frac {K^4 \tau^2} {2
\sigma^2}} \sum \limits_{m \neq 0} {\frac {m^2} {\omega_m^2}}
{\left( {\frac {\partial^2 A_m} {\partial \xi^2}} e^{i z_m^{(+)}}
+ {\frac {\partial^2 B_m} {\partial \xi^2}} e^{i z_m^{(-)}}
\right)} \nonumber
\end{eqnarray}
\begin{eqnarray}
+ {\frac {K^2 \tau} {2i \sigma^2}} \sum \limits_{m \neq 0} {\frac
{1} {\omega_m^3}} {\left( {\frac {\partial^2 A_m} {\partial
\xi^2}} e^{i z_m^{(+)}} + {\frac {\partial^2 B_m} {\partial
\xi^2}} e^{i z_m^{(-)}} \right)} \nonumber
\end{eqnarray}
\begin{equation}
+ {\frac {2 \pi^2 \tau} {i \sigma^2}} \sum \limits_{m, n \neq 0}
{\frac {1} {\omega_m}} {\left[ {\left( \Gamma_{mn}^{(+)} {\left|
A_n \right|}^2 + \Gamma_{mn}^{(-)} {\left| B_n \right|}^2 \right)}
A_m e^{i z_m^{(+)}} + {\left( \Gamma_{mn}^{(-)} {\left| A_n
\right|}^2 + \Gamma_{mn}^{(+)} {\left| B_n \right|}^2 \right)} B_m
e^{i z_m^{(-)}} \right]}, \label{Solr2}
\end{equation}
\noindent where non-secular oscillating terms have not been
written out in full explicitly. The final step is to collect the
terms proportional to the fundamental modes $e^{i z_m^{(+)}}$ and
$e^{i z_m^{(-)}}$ in all orders and renormalize the amplitudes
$A_m$ and $B_m$. As a result one obtains the following RG
equations
\begin{equation}
2i \omega_m {\frac {\partial A_m} {\partial \tau}} - 2im {\frac
{K^2} {\sigma}} {\frac {\partial A_m} {\partial x}} = {\frac {K^2}
{\sigma^2 \omega_m^2}} {\frac {\partial^2 A_m} {\partial x^2}} +
{\frac {4 \pi^2} {\sigma^2}} A_m \sum \limits_{n \neq 0} {\left(
\Gamma_{mn}^{(+)} {\left| A_n \right|}^2 + \Gamma_{mn}^{(-)}
{\left| B_n \right|}^2 \right)}, \label{RGequata}
\end{equation}
\begin{equation}
2i \omega_m {\frac {\partial B_m} {\partial \tau}} + 2im {\frac
{K^2} {\sigma}} {\frac {\partial B_m} {\partial x}} = {\frac {K^2}
{\sigma^2 \omega_m^2}} {\frac {\partial^2 B_m} {\partial x^2}} +
{\frac {4 \pi^2} {\sigma^2}} B_m \sum \limits_{n \neq 0} {\left(
\Gamma_{mn}^{(-)} {\left| A_n \right|}^2 + \Gamma_{mn}^{(+)}
{\left| B_n \right|}^2 \right)}, \label{RGequatb}
\end{equation}

\renewcommand{\theequation}{\thesection.\arabic{equation}}

\setcounter{equation}{0}

\section{The Parametric Resonance}

Let us now address the system of equations (\ref{Basic1}) and
(\ref{Basic2}). Without loss of generality, we assume that the
time variation of $g(\tau)$ can be treated as a second-order
perturbation (which is usually the case), that is
\begin{equation}
g(\tau) = a_0 + \epsilon^2 \sum \limits_{n \neq 0} a_n e^{in \tau
/ \nu}. \label{Fourier}
\end{equation}
\noindent The same holds true for the envelope function ${\cal E}
(\tau)$
\begin{equation}
{\cal E} (\tau) = \lambda a_0 + \epsilon^2 \lambda \sum \limits_{n
\neq 0} b_n e^{in \tau / \nu}, \qquad \qquad b_n = {\frac {a_n} {1
- n^2 / \nu^2}}. \label{Fourier1}
\end{equation}
\noindent Thus, the basic equations (\ref{Basic1}) and
(\ref{Basic2}) can be rewritten in the form
\begin{eqnarray}
{\cal E} {\frac {\partial} {\partial \tau}} {\left( {\cal E}
{\frac {\partial R} {\partial \tau}} \right)} - 2 v_T^2 {\frac
{\partial^2 R} {\partial x^2}} + \lambda g {\cal E} R \nonumber
\end{eqnarray}
\begin{equation}
= \epsilon^2 {\dot{\cal E}}_2 {\frac {\partial^2} {\partial x^2}}
(xu) - \epsilon^2 {\cal E} {\frac {\partial} {\partial \tau}}
{\left[ {\dot{\cal E}}_2 {\frac {\partial} {\partial x}} (xR)
\right]} + \epsilon {\cal E} {\frac {\partial^2} {\partial x^2}}
{\left( {\frac {u^2} {2}} + v_T^2 R^2 \right)} - \epsilon {\cal E}
{\frac {\partial} {\partial \tau}} {\left[ {\cal E} {\frac
{\partial} {\partial x}} (Ru) \right]}. \label{Basic1pr}
\end{equation}
\begin{equation}
{\cal E} {\frac {\partial R} {\partial \tau}} + {\frac {\partial
u} {\partial x}} + \epsilon^2 {\dot{\cal E}}_2 {\frac {\partial}
{\partial x}} (xR) + \epsilon {\cal E} {\frac {\partial} {\partial
x}} {\left( R u \right)} = 0, \label{Basic2pr}
\end{equation}
\noindent Let us reiterate that the assumption concerning the
second order of magnitude in $\epsilon$ of the time variation of
$g (\tau)$ and ${\cal E} (\tau)$ does not restrict the generality
of the subsequent analysis. If this variation were of first order
in $\epsilon$, the proper perturbation parameter to use would be
$\sqrt{\epsilon}$ instead of $\epsilon$. In addition, the
variables $R$ and $u$ have to be rescaled accordingly
\begin{equation}
R \longrightarrow {\frac {R} {\sqrt{\epsilon}}}, \qquad \qquad u
\longrightarrow {\frac {u} {\sqrt{\epsilon}}}. \label{Rescale}
\end{equation}
\noindent In terms of the ansatz (\ref{Actdepen}) and
(\ref{Xivar}), equations (\ref{Basic1pr}) and (\ref{Basic2pr})
become
\begin{eqnarray}
{\cal E} {\frac {\partial} {\partial \tau}} {\left( {\cal E}
{\frac {\partial R} {\partial \tau}} \right)} - 2 v_T^2 {\left(
{\frac {\partial^2} {\partial x^2}} + 2 \epsilon {\frac
{\partial^2} {\partial x \partial \xi}} + \epsilon^2 {\frac
{\partial^2} {\partial \xi^2}} \right)} R + \lambda g {\cal E} R
\nonumber
\end{eqnarray}
\begin{eqnarray}
= \epsilon {\dot{\cal E}}_2 {\left( {\frac {\partial^2} {\partial
x^2}} + 2 \epsilon {\frac {\partial^2} {\partial x \partial \xi}}
+ \epsilon^2 {\frac {\partial^2} {\partial \xi^2}} \right)}
{\left( \xi u \right)} - \epsilon {\cal E} {\frac {\partial}
{\partial \tau}} {\left[ {\dot{\cal E}}_2 {\left( {\frac
{\partial} {\partial x}} + \epsilon {\frac {\partial} {\partial
\xi}} \right)} {\left( \xi R \right)} \right]} \nonumber
\end{eqnarray}
\begin{equation}
+ \epsilon {\cal E} {\left( {\frac {\partial^2} {\partial x^2}} +
2 \epsilon {\frac {\partial^2} {\partial x \partial \xi}} +
\epsilon^2 {\frac {\partial^2} {\partial \xi^2}} \right)} {\left(
{\frac {u^2} {2}} + v_T^2 R^2 \right)} - \epsilon {\cal E} {\frac
{\partial} {\partial \tau}} {\left[ {\cal E} {\left( {\frac
{\partial} {\partial x}} + \epsilon {\frac {\partial} {\partial
\xi}} \right)} {\left( R u \right)} \right]}, \label{Basic12pr}
\end{equation}
\begin{equation}
{\cal E} {\frac {\partial R} {\partial \tau}} + {\frac {\partial
u} {\partial x}} + \epsilon {\frac {\partial u} {\partial \xi}} +
\epsilon {\dot{\cal E}}_2 {\left( {\frac {\partial} {\partial x}}
+ \epsilon {\frac {\partial} {\partial \xi}} \right)} {\left( \xi
R \right)} + \epsilon {\cal E} {\left( {\frac {\partial} {\partial
x}} + \epsilon {\frac {\partial} {\partial \xi}} \right)} {\left(
R u \right)} = 0. \label{Basic21pr}
\end{equation}
\noindent Although the general case can be in principle treated
through more labour-intensive manipulations, for the sake of
simplicity in what follows, we select a particular mode with mode
number $m$. The zeroth-order solution can be written in the form
\begin{equation}
R_0 {\left( x, \xi; \tau \right)} = A {\left( \xi \right)} e^{i
z^{(+)} {\left( x, \tau \right)}} + B {\left( \xi \right)} e^{i
z^{(-)} {\left( x, \tau \right)}} + c.c., \label{Zeroordrpr}
\end{equation}
\begin{equation}
u_0 {\left( x, \xi; \tau \right)} = - {\frac {2 \pi \omega} {m
\sigma^2}} {\left[ A {\left( \xi \right)} e^{i z^{(+)} {\left( x,
\tau \right)}} - B {\left( \xi \right)} e^{i z^{(-)} {\left( x,
\tau \right)}} \right]} + c.c., \label{Zeroordupr}
\end{equation}
\noindent where again
\begin{eqnarray}
z^{(\pm)} {\left( x, \tau \right)} = \omega \tau \pm m \sigma x,
\qquad \qquad \omega^2 = 1 + K^2 m^2. \nonumber
\end{eqnarray}
\noindent The particular mode is chosen such that an exact
parametric resonance (if possible)
\begin{equation}
2 \omega = {\frac {n_0} {\nu}} \label{Parametric}
\end{equation}
\noindent occurs for some integer $n_0$. The first order
perturbation equation for $R_1$ can be written in the form
\begin{eqnarray}
{\frac {\partial^2 R_1} {\partial \tau^2}} - {\frac {K^2}
{\sigma^2}} {\frac {\partial^2 R_1} {\partial x^2}} + R_1 = {\frac
{2imK^2} {\sigma}} {\left( {\frac {\partial A} {\partial \xi}}
e^{i z^{(+)}} - {\frac {\partial B} {\partial \xi}} e^{i z^{(-)}}
\right)} \nonumber
\end{eqnarray}
\begin{eqnarray}
+ {\frac {i m \sigma^2} {2 \pi}} \xi \lambda {\sum \limits_{n \neq
0}}^{\prime} {\frac {n} {\nu}} {\left( 2 \omega + {\frac {n}
{\nu}} \right)} b_n e^{in \tau / \nu} {\left( A e^{i z^{(+)}} - B
e^{i z^{(-)}} \right)} \nonumber
\end{eqnarray}
\begin{equation}
- {\frac {4 \pi} {\sigma}} {\left( 4 \omega^2 - 1 \right)} {\left(
A^2 e^{2i z^{(+)}} + B^2 e^{2i z^{(-)}} \right)} + {\frac {8 \pi}
{\sigma}} AB^{\ast} e^{i {\left( z^{(+)} - z^{(-)} \right)}} +
c.c.. \label{Firstrpr}
\end{equation}
\noindent Here ${\sum}^{\prime}$ implies exclusion of the harmonic
with $n = n_0$ from the sum. It is important to note that terms
giving rise to secular contribution due to the parametric
resonance vanish identically in the first order. The solution for
$R_1$ can be written explicitly as
\begin{eqnarray}
R_1 {\left( x, \xi; \tau \right)} = {\frac {mK^2} {\sigma \omega}}
\tau {\left( {\frac {\partial A} {\partial \xi}} e^{i z^{(+)}} -
{\frac {\partial B} {\partial \xi}} e^{i z^{(-)}} \right)} -
{\frac {i m \sigma^2} {2 \pi}} \xi \lambda {\sum \limits_{n \neq
0}}^{\prime} b_n e^{in \tau / \nu} {\left( A e^{i z^{(+)}} - B
e^{i z^{(-)}} \right)} \nonumber
\end{eqnarray}
\begin{equation}
+ {\frac {4 \pi} {3 \sigma}} {\left( 4 \omega^2 - 1 \right)}
{\left( A^2 e^{2i z^{(+)}} + B^2 e^{2i z^{(-)}} \right)} + {\frac
{8 \pi} {\sigma {\left( 4 \omega^2 - 3 \right)}}} AB^{\ast} e^{i
{\left( z^{(+)} - z^{(-)} \right)}} + c.c.. \label{Solfirstrpr}
\end{equation}
\noindent Straightforward calculations yield the solution for
$u_1$
\begin{eqnarray}
u_1 {\left( x, \xi; \tau \right)} = -  {\frac {2 \pi i} {m^2
\sigma^3 \omega}} {\left( {\frac {\partial A} {\partial \xi}} e^{i
z^{(+)}} + {\frac {\partial B} {\partial \xi}} e^{i z^{(-)}}
\right)} - {\frac {2 \pi K^2} {\sigma^3}} \tau {\left( {\frac
{\partial A} {\partial \xi}} e^{i z^{(+)}} + {\frac {\partial B}
{\partial \xi}} e^{i z^{(-)}} \right)} \nonumber
\end{eqnarray}
\begin{eqnarray}
- {\frac {i \lambda n_0} {\nu}} \xi {\left( b_{n_0} e^{in_0 \tau /
\nu} - b_{n_0}^{\ast} e^{-in_0 \tau / \nu}  \right)} {\left( A
e^{i z^{(+)}} + B e^{i z^{(-)}} \right)} \nonumber
\end{eqnarray}
\begin{equation}
+ i \omega \xi \lambda {\sum \limits_{n \neq 0}}^{\prime} b_n
e^{in \tau / \nu} {\left( A e^{i z^{(+)}} + B e^{i z^{(-)}}
\right)} - {\frac {8 \pi^2} {3m \sigma^3}} \omega {\left( 4
\omega^2 - 5/2 \right)} {\left( A^2 e^{2i z^{(+)}} - B^2 e^{2i
z^{(-)}} \right)} + c.c.. \label{Solfirstupr}
\end{equation}

In second order, we take into account terms providing secular
contribution to the solution only. Thus, we write the equation for
$R_2$ as
\begin{eqnarray}
{\frac {\partial^2 R_2} {\partial \tau^2}} - {\frac {K^2}
{\sigma^2}} {\frac {\partial^2 R_2} {\partial x^2}} + R_2 = {\frac
{2iK^4 m^2} {\omega \sigma^2}} \tau {\left( {\frac {\partial^2 A}
{\partial \xi^2}} e^{i z^{(+)}} + {\frac {\partial^2 B} {\partial
\xi^2}} e^{i z^{(-)}} \right)} \nonumber
\end{eqnarray}
\begin{eqnarray}
+ {\frac {\lambda \sigma} {\pi}} b_{n_0} {\left( 1 - \omega^2
\right)} {\left[ {\left( \xi {\frac {\partial} {\partial \xi}} - 1
\right)} B^{\ast} e^{i z^{(+)}} + {\left( \xi {\frac {\partial}
{\partial \xi}} - 1 \right)} A^{\ast} e^{i z^{(-)}} \right]}
\nonumber
\end{eqnarray}
\begin{eqnarray}
+ {\frac {K^2} {\sigma^2}} {\left( {\frac {\partial^2 A} {\partial
\xi^2}} e^{i z^{(+)}} + {\frac {\partial^2 B} {\partial \xi^2}}
e^{i z^{(-)}} \right)} + {\frac {2 \lambda^2 \sigma^2} {\pi^2}}
\omega^2 m^2 \sigma^2 {\left| b_{n_0} \right|}^2 \xi^2 {\left( A
e^{i z^{(+)}} + B e^{i z^{(-)}} \right)} \nonumber
\end{eqnarray}
\begin{eqnarray}
- {\frac {8 \pi^2} {3 \sigma^2}} {\left( 16 \omega^4 - 11 \omega^2
+ 1 \right)} {\left( {\left| A \right|}^2 A e^{i z^{(+)}} +
{\left| B \right|}^2 B e^{i z^{(-)}} \right)} \nonumber
\end{eqnarray}
\begin{equation}
+ {\frac {16 \pi^2} {\sigma^2 {\left( 4 \omega^2 - 3 \right)}}}
{\left( {\left| B \right|}^2 A e^{i z^{(+)}} + {\left| A
\right|}^2 B e^{i z^{(-)}} \right)} + c.c.. \label{Secondpr}
\end{equation}
\noindent Similar to equation (\ref{Solr2}), we obtain in a
straightforward manner
\begin{eqnarray}
R_2 {\left( x, \xi; \tau \right)} = {\frac {K^4 m^2} {2 \omega^2
\sigma^2}} \tau^2 {\left( {\frac {\partial^2 A} {\partial \xi^2}}
e^{i z^{(+)}} + {\frac {\partial^2 B} {\partial \xi^2}} e^{i
z^{(-)}} \right)} + {\frac {K^2 \tau} {2i \omega^3 \sigma^2}}
{\left( {\frac {\partial^2 A} {\partial \xi^2}} e^{i z^{(+)}} +
{\frac {\partial^2 B} {\partial \xi^2}} e^{i z^{(-)}} \right)}
\nonumber
\end{eqnarray}
\begin{eqnarray}
+ {\frac {\lambda \sigma \tau} {2 \pi i \omega}} b_{n_0} {\left( 1
- \omega^2 \right)} {\left[ {\left( \xi {\frac {\partial}
{\partial \xi}} - 1 \right)} B^{\ast} e^{i z^{(+)}} + {\left( \xi
{\frac {\partial} {\partial \xi}} - 1 \right)} A^{\ast} e^{i
z^{(-)}} \right]} \nonumber
\end{eqnarray}
\begin{eqnarray}
+ {\frac {2 \lambda^2 \sigma^2 \tau} {2i \pi^2 \omega}} \omega^2
m^2 \sigma^2 {\left| b_{n_0} \right|}^2 \xi^2 {\left( A e^{i
z^{(+)}} + B e^{i z^{(-)}} \right)} \nonumber
\end{eqnarray}
\begin{eqnarray}
- {\frac {\tau} {2i \omega}} {\frac {8 \pi^2} {3 \sigma^2}}
{\left( 16 \omega^4 - 11 \omega^2 + 1 \right)} {\left( {\left| A
\right|}^2 A e^{i z^{(+)}} + {\left| B \right|}^2 B e^{i z^{(-)}}
\right)} \nonumber
\end{eqnarray}
\begin{equation}
+ {\frac {\tau} {2i \omega}} {\frac {16 \pi^2} {\sigma^2 {\left( 4
\omega^2 - 3 \right)}}} {\left( {\left| B \right|}^2 A e^{i
z^{(+)}} + {\left| A \right|}^2 B e^{i z^{(-)}} \right)} + c.c..
\label{Secsolpr}
\end{equation}
\noindent Collecting once again the zeroth-order term together
with all secular terms in higher orders and renormalizing the
amplitudes $A$ and $B$, we obtain the RG equations
\begin{eqnarray}
2i \omega {\frac {\partial A} {\partial \tau}} - 2im {\frac {K^2}
{\sigma}} {\frac {\partial A} {\partial x}} = {\frac {K^2}
{\sigma^2 \omega^2}} {\frac {\partial^2 A} {\partial x^2}} +
{\frac {\lambda \sigma} {\pi}} b_{n_0} {\left( 1 - \omega^2
\right)} {\left( x {\frac {\partial} {\partial x}} - 1 \right)}
B^{\ast} \nonumber
\end{eqnarray}
\begin{equation}
+ {\frac {2 \lambda^2 \sigma^4} {\pi^2}} m^2 \omega^2 {\left|
b_{n_0} \right|}^2 x^2 A - {\frac {8 \pi^2} {3 \sigma^2}} {\left(
16 \omega^4 - 11 \omega^2 + 1 \right)} {\left| A \right|}^2 A +
{\frac {16 \pi^2} {\sigma^2 {\left( 4 \omega^2 - 3 \right)}}}
{\left| B \right|}^2 A, \label{RGeqapr}
\end{equation}
\begin{eqnarray}
2i \omega {\frac {\partial B} {\partial \tau}} + 2im {\frac {K^2}
{\sigma}} {\frac {\partial B} {\partial x}} = {\frac {K^2}
{\sigma^2 \omega^2}} {\frac {\partial^2 B} {\partial x^2}} +
{\frac {\lambda \sigma} {\pi}} b_{n_0} {\left( 1 - \omega^2
\right)} {\left( x {\frac {\partial} {\partial x}} - 1 \right)}
A^{\ast} \nonumber
\end{eqnarray}
\begin{equation}
+ {\frac {2 \lambda^2 \sigma^4} {\pi^2}} m^2 \omega^2 {\left|
b_{n_0} \right|}^2 x^2 B - {\frac {8 \pi^2} {3 \sigma^2}} {\left(
16 \omega^4 - 11 \omega^2 + 1 \right)} {\left| B \right|}^2 B +
{\frac {16 \pi^2} {\sigma^2 {\left( 4 \omega^2 - 3 \right)}}}
{\left| A \right|}^2 B. \label{RGeqbpr}
\end{equation}

\renewcommand{\theequation}{\thesection.\arabic{equation}}

\setcounter{equation}{0}

\section{The Nonlinear Schrodinger Equation for a Single Mode}

Equations (\ref{RGequata}) and (\ref{RGequatb}) represent a system
of coupled nonlinear Schrodinger equations for the mode amplitudes
$A_m$ and $B_m$. Neglecting the contribution from modes with $n
\neq \pm m$ and introducing a new amplitude ${\widetilde{B}}_m
{\left( x; \tau \right)} = B_m {\left( -x; \tau \right)}$, for the
single mode amplitudes $A_m$ and $B_m$, we obtain the equations
\begin{equation}
i {\frac {\partial A} {\partial \tau}} - {\frac {\partial^2 A}
{\partial \zeta^2}} - {\frac {8 \pi^2} {\omega \sigma^2 {\left( 4
\omega^2 - 3 \right)}}} {\left( -G {\left| A \right|}^2 + {\left|
B \right|}^2 \right)} A = 0, \label{RGeqasmode}
\end{equation}
\begin{equation}
i {\frac {\partial B} {\partial \tau}} - {\frac {\partial^2 B}
{\partial \zeta^2}} - {\frac {8 \pi^2} {\omega \sigma^2 {\left( 4
\omega^2 - 3 \right)}}} {\left( {\left| A \right|}^2  -G {\left| B
\right|}^2 \right)} B = 0, \label{RGeqbsmode}
\end{equation}
\noindent where
\begin{equation}
\zeta = {\sqrt{2 \omega}} {\left( mK \tau + {\frac {\omega \sigma}
{K}} \right)}, \qquad \qquad G = {\frac {4 \omega^2 - 3} {6}}
{\left( 16 \omega^4 - 11 \omega^2 + 1 \right)}, \label{Varcoeff}
\end{equation}
\noindent the index $m$ and the tilde sign over the new amplitude
${\widetilde{B}}$ has been omitted. Clearly enough, equations
(\ref{RGeqasmode}) and (\ref{RGeqbsmode}) follow directly from the
system (\ref{RGeqapr}) - (\ref{RGeqbpr}) if $b_{n_0}$ is set to
zero. This implies that in the case where a parametric resonance
does not occur in the original system, the smooth approximation is
valid to second order in the formal perturbation parameter.
Moreover, it is straightforward to verify that $G$ is always
positive. The system of two coupled nonlinear Schrodinger
equations (\ref{RGeqasmode}) and (\ref{RGeqbsmode}) is in general
non integrable. Its integrability is proven by Manakov
\cite{manakov} only in the simplest case of $G = -1$.

If one of the amplitudes ($B$ or $A$) in its capacity of being a
particular solution to the corresponding nonlinear Schrodinger
equation is identically zero, the equation for the other amplitude
(say $A$) becomes
\begin{equation}
i {\frac {\partial A} {\partial \tau}} - {\frac {\partial^2 A}
{\partial \zeta^2}} + \Gamma {\left| A \right|}^2 A = 0,
\label{Nlsedark}
\end{equation}
\noindent where
\begin{equation}
\Gamma = {\frac {4 \pi^2} {3 \omega \sigma^2}} {\left( 16 \omega^4
- 11 \omega^2 + 1 \right)}. \label{Gammacoef}
\end{equation}
\noindent In nonlinear optics equation (\ref{Nlsedark}) is known
to describe the formation and evolution of the so-called {\it dark
solitons} \cite{kivshar}. In the case of charged particle beams
these correspond to the formation of {\it holes} or {\it cavitons}
in the beam. The solution to equation (\ref{Nlsedark}) can be
written as
\begin{equation}
A {\left( \zeta; \tau \right)} = {\frac {r {\left( \xi \right)}}
{\sqrt{\Gamma}}} e^{i {\left[ n \tau + \theta {\left( \xi \right)}
\right]}}, \qquad \qquad \xi = \zeta - c \tau, \label{Nlsadasol}
\end{equation}
\noindent where
\begin{equation}
r {\left( \xi \right)} = {\sqrt{n - 2 a^2 {\rm sech}^2 {\left( a
\xi \right)}}}, \qquad \qquad \theta {\left( \xi \right)} =
\arctan {\left[ {\frac {2a} {c}} {\rm tanh} {\left( a \xi \right)}
\right]}, \label{Rthetafun}
\end{equation}
\noindent for all $c$ and
\begin{equation}
a = {\frac {1} {2}} {\sqrt{2n - c^2}}, \label{Coeffa}
\end{equation}
\noindent provided $n > c^2 / 2$.

Let us now examine equations (\ref{RGeqapr}) and (\ref{RGeqbpr}).
In the cold-beam limit $v_T \rightarrow 0$ (or equivalently
$\omega \rightarrow 1$) the second term on their right-hand-sides
can be neglected as compared to the other terms. Therefore, we can
write
\begin{eqnarray}
2i \omega {\frac {\partial {\widetilde{A}}} {\partial \tau}} =
{\frac {K^2} {\sigma^2 \omega^2}} {\frac {\partial^2
{\widetilde{A}}} {\partial x^2}} + {\frac {2 \lambda^2 \sigma^4}
{\pi^2}} m^2 \omega^2 {\left| b_{n_0} \right|}^2 x^2
{\widetilde{A}} \nonumber
\end{eqnarray}
\begin{equation}
- {\frac {8 \pi^2} {3 \sigma^2}} {\left( 16 \omega^4 - 11 \omega^2
+ 1 \right)} {\left| {\widetilde{A}} \right|}^2 {\widetilde{A}} +
{\frac {16 \pi^2} {\sigma^2 {\left( 4 \omega^2 - 3 \right)}}}
{\left| {\widetilde{B}} \right|}^2 {\widetilde{A}},
\label{RGeqagp}
\end{equation}
\begin{eqnarray}
2i \omega {\frac {\partial {\widetilde{B}}} {\partial \tau}} =
{\frac {K^2} {\sigma^2 \omega^2}} {\frac {\partial^2
{\widetilde{B}}} {\partial x^2}} + {\frac {2 \lambda^2 \sigma^4}
{\pi^2}} m^2 \omega^2 {\left| b_{n_0} \right|}^2 x^2
{\widetilde{B}} \nonumber
\end{eqnarray}
\begin{equation}
- {\frac {8 \pi^2} {3 \sigma^2}} {\left( 16 \omega^4 - 11 \omega^2
+ 1 \right)} {\left| {\widetilde{B}} \right|}^2 {\widetilde{B}} +
{\frac {16 \pi^2} {\sigma^2 {\left( 4 \omega^2 - 3 \right)}}}
{\left| {\widetilde{A}} \right|}^2 {\widetilde{B}},
\label{RGeqbgp}
\end{equation}
\noindent where the local gauge transformation
\begin{equation}
{\widetilde{A}} = A \exp {\left[ i m \omega {\left( {\frac {m K^2}
{2}} \tau + \sigma \omega x \right)} \right]}, \qquad \qquad
{\widetilde{B}} = B \exp {\left[ i m \omega {\left( {\frac {m K^2}
{2}} \tau - \sigma \omega x \right)} \right]}, \label{Localgauge}
\end{equation}
\noindent has been performed on the amplitudes $A$ and $B$.
Equations (\ref{RGeqagp}) and (\ref{RGeqbgp}) represent a system
of two coupled Gross-Pitaevskii equations, known to govern the
formation of condensate structures in atomic gases confined in
magnetic traps. In addition, superfluidity in helium as well as
patterns in the gas of paraexcitons in semiconductors are
considered as a possible manifestation of Bose-Einstein
condensation. It is remarkable that under certain conditions
similar phenomenon can be observed in space-charge dominated
beams.

\section{Concluding Remarks}

The analysis performed in the present paper is based on the
Vlasov-Maxwell equations for the self-consistent evolution of the
beam distribution function and the electromagnetic fields. We
considered the propagation of an intense beam through a periodic
focusing lattice. It has been proven that a special class of
solutions to the Vlasov-Maxwell equations exists, which defines a
uniform phase-space density inside of simply connected boundary
curves. These solutions provide an exact closure to the hierarchy
of moments resulting in a set of hydrodynamic equations for the
beam density (continuity equation) and for the current velocity.
The latter is characterized by a triple adiabatic pressure law.
The major drawback of the model discussed here is that due to the
constancy of the distribution function inside of the evolving
region of phase space, it fails to take into account an important
feature such as the well-known Landau damping. Inclusion of the
Landau damping in the description can be achieved correctly only
by direct analysis of the Vlasov-Maxwell system \cite{tzenov}.

Further, we studied first the case when the smooth focusing
approximation applies. Based on the RG method, a system of coupled
nonlinear Schrodinger equations has been derived for the slowly
varying amplitudes of interacting beam-density waves. Under the
approximation of an isolated mode neglecting the effect of the
rest of the modes, this system reduces to two coupled nonlinear
Schrodinger equations for the amplitudes of the forward and of the
backward wave. The particular solution to the latter system
asserting that the amplitude of either of the waves (the forward
or the backward) can vanish identically, leads to a single
nonlinear Schrodinger equation for the other amplitude. The latter
is characterized by a repulsive nonlinearity and therefore, it
describes the evolution of {\it holes} in intense charged particle
beams.

The analysis of periodic focusing clearly showed that the results
obtained in the case of smooth focusing remain unchanged up to
second order in the perturbation parameter if a parametric
resonance between a particular mode and an appropriate Fourier
harmonic of the external focusing does not occur. If however, an
exact parametric resonance takes place, it was shown that the
evolution of the wave amplitudes of the resonant mode in the
cold-beam limit is described by a system of coupled
Gross-Pitaevskii equations. Quite remarkably, it was found that
there exist a possibility of formation of density condensates in
space-charge dominated beams.

\subsection*{Acknowledgments}

It is a pleasure to thank Y. Oono and S.-I. Goto for careful
reading of the manuscript and for making valuable suggestions. I
am also greatly indebted to B. Baizakov and F. Illuminati for many
helpful discussion concerning the subject of the present paper and
for their valuable comments on its early version. This research
was supported by the INFM grant number OA03002735.

\renewcommand{\theequation}{A.\arabic{equation}}

\setcounter{equation}{0}

\section*{Appendix}

In order to prove that the expression (\ref{Dfunexform})
represents the solution of the Vlasov equation (\ref{Vlasoveq}),
we write the equations for the characteristics of the latter in
the form
\begin{equation}
{\frac {{\rm d} \theta} {1}} = {\frac {{\rm d} x} {{\dot{\chi}}
p}} = - {\rm d} p {\left( {\dot{\chi}} x + {\frac {\partial V}
{\partial x}} + \lambda {\sqrt{\beta}} {\frac {\partial U}
{\partial x}} \right)}^{-1} = {\frac {{\rm d} f} {0}}.
\label{Characteq}
\end{equation}
\noindent Let us also assume that the distribution function $f
{\left( x, p; \theta_0 \right)}$ at some initial time $\theta_0$
is given by
\begin{equation}
f {\left( x, p; \theta_0 \right)} = f_0 {\left( x, p \right)} =
{\cal C} {\left[ {\cal H} {\left( p - p_{(-)}^{(0)} {\left( x
\right)} \right)} - {\cal H} {\left( p - p_{(+)}^{(0)} {\left( x
\right)} \right)} \right]}. \label{Initdistr}
\end{equation}
\noindent Suppose now that we have been able to solve the
equations for the characteristics (\ref{Characteq}), and we have
expressed the solution subsequently according to
\begin{equation}
p = P {\left( x; \theta \right)}, \label{Equivexp}
\end{equation}
\noindent where $P {\left( x; \theta \right)}$ is an appropriate
function of its arguments. At each instant of time $\theta$
equation (\ref{Equivexp}) defines two curves $p_{(+)} {\left( x;
\theta \right)}$ and $p_{(-)} {\left( x; \theta \right)}$ in the
phase space $(x, p)$, such that
\begin{equation}
p_{(+)} {\left( x; \theta_0 \right)} = p_{(+)}^{(0)} {\left( x
\right)}, \qquad \qquad p_{(-)} {\left( x; \theta_0 \right)} =
p_{(-)}^{(0)} {\left( x \right)}. \label{Twocurves}
\end{equation}
\noindent Therefore, if initially the distribution function is
given by equation (\ref{Initdistr}), then the solution of the
Vlasov equation (\ref{Vlasoveq}) at every subsequent instant of
time $\theta$ is represented by the expression (\ref{Dfunexform}).

It is interesting to note that the chain of equations (excluding
the last equation) for the characteristics (\ref{Characteq}) of
the Vlasov equation formally coincide with those for the equation
\begin{equation}
{\frac {\partial P} {\partial \theta}} + {\dot{\chi}} P {\frac
{\partial P} {\partial x}} = - {\dot{\chi}} x - {\frac {\partial
V} {\partial x}} - \lambda {\sqrt{\beta}} {\frac {\partial U}
{\partial x}}. \label{Phydrodyn}
\end{equation}
\noindent As a matter of fact, equation (\ref{Phydrodyn}) is
equivalent to the two equations
\begin{equation}
{\frac {\partial p_{(\pm)}} {\partial \theta}} + {\dot{\chi}}
p_{(\pm)} {\frac {\partial p_{(\pm)}} {\partial x}} = -
{\dot{\chi}} x - {\frac {\partial V} {\partial x}} - \lambda
{\sqrt{\beta}} {\frac {\partial U} {\partial x}},
\label{Ppmhydrodyn}
\end{equation}
\noindent with the initial conditions (\ref{Twocurves}), provided
$U$ is defined as a solution to the equation (\ref{Poissonmom}).

%%%%% Bibliography
%%


\begin{thebibliography}{99}

\bibitem{prstab} R.C. Davidson, H. Qin, S.I. Tzenov and E.A.
Startsev, {\it Phys. Rev. ST Accel. Beams} {\bf 5}, 084402 (2002).

\bibitem{cgoono} L.Y. Chen, N. Goldenfeld and Y. Oono, {\it Phys.
Rev.} {\bf E} {\bf 54}, 376 (1996).

\bibitem{nozaki} K. Nozaki, Y. Oono and Y. Shiwa, {\it Phys.
Rev.} {\bf E} {\bf 62}, 4501 (2000).

\bibitem{oono} K. Nozaki and Y. Oono, {\it Phys. Rev.} {\bf E} {\bf 63},
046101 (2001).

\bibitem{shiwa} Y. Shiwa, {\it Phys. Rev.} {\bf E} {\bf 63}, 016119
(2001).

\bibitem{dalfovo} F. Dalfovo, S. Giorgini, L.P. Pitaevskii and S.
Stringari, {\it Rev. Mod. Phys.} {\bf 71}, 463-512 (1999).

\bibitem{nayfeh} A.H. Nayfeh, ``{\it Introduction to Perturbation
Techniques}'' (Wiley, New York, 1981).

\bibitem{manakov} S.V. Manakov, ``{\it On the Theory of Two-Dimensional
Stationary Self-Focusing of Electromagnetic Waves}'' {\it Zh.
Eksp. Teor. Fiz. [Sov. Phys. JETP]} {\bf 65 [38]}, 505-516
[248-253] (1973 [1974]).

\bibitem{kivshar} Y.S. Kivshar and B. Luther-Davies, {\it Phys. Rep.}
{\bf 298}, 81 (1998).

\bibitem{tzenov} S.I. Tzenov, "{\it Nonlinear Landau Damping in Space-Charge
Dominated Beams}" In preparation to be submitted for publication.

\end{thebibliography}
\end{document}